\documentclass[prd,amsmath,amssymb,longbibliography,superscriptaddress,twocolumn,nofootinbib,10pt]{revtex4}

\pdfoutput=1

\usepackage{graphicx}
\usepackage{dcolumn}
\usepackage{bm}
\usepackage{amssymb}
\usepackage{latexsym}
\usepackage{booktabs}
\usepackage{amsmath}
\usepackage{multirow}
\usepackage[colorlinks=true, linkcolor=blue, citecolor=blue]{hyperref}

\def\aap{Astron. Astrophys.}
\def\mnras{Mon. Not. Roy. Astron. Soc.}
\def\apjl{Astrophys. J. Lett.}
\def\jcap{JCAP}
\def\nar{New Astronomy Reviews}

\newcommand{\LCDM}{\rm{\Lambda}CDM}

\newcommand{\Mpc}{\mathrm{~km~s^{-1}~Mpc^{-1}}}
\begin{document}

\title{A model-independent determination of the sound horizon using recent BAO measurements and strong lensing systems}

\author{Tonghua Liu}
\affiliation{School of Physics and Optoelectronic, Yangtze University, Jingzhou 434023, China;}
\author{Shuo Cao}
\email{caoshuo@bnu.edu.cn}
\affiliation{School of Physics and Astronomy, Beijing Normal University, Beijing 100875, China;}
\affiliation{Institute for Frontiers in Astronomy and Astrophysics, Beijing Normal University, Beijing 102206, China;}
\author{Jieci Wang}
\email{jcwang@hunnu.edu.cn}
\affiliation{Department of Physics, and Collaborative Innovation Center for Quantum Effects and Applications, Hunan Normal University, Changsha 410081, China;}

\begin{abstract}
We propose an improved method to determine the sound horizon in a cosmological model-independent way by using the latest observations of BAO measurements from DES, BOSS/eBOSS, and DESI surveys and gravitationally time-delay lensed quasars from H0LiCOW collaboration.
Combining the 6$D_{\Delta t}$ plus 4$D_{d}$ measurements and the reconstructed BAO datasets, we obtain a model-independent result of $r_d=139.7^{+5.2}_{-4.5}$ Mpc, with the precision at the $\sim3.7\%$ level, which is in agreement with the result of Planck 2018 within $\sim1.7\sigma$ uncertainty.
Our method is independent of cosmological parameters such as the Hubble constant, dark energy, (and, more importantly, does not involve the cosmic curvature when using the $D_d$ measurements of the lenses, and also avoids the obstacle of mass-sheet degeneracy in gravitational lensing). Meanwhile, it does not need to consider the Eddington relation with concerning the transformation of distance. Since only two types of data are considered, the contribution of each can be clearly understood. Our results also highlight the Hubble tension and may give us a better understanding of the discordance between the datasets or reveal new physics beyond the standard model.

\end{abstract}

\maketitle

\section{Introduction}
In the early universe, baryon acoustic oscillation (BAOs) are sound waves produced by gravitational interactions between photon-baryon fluids and inhomogeneity \cite{1984ucp..book.....W,1972gcpa.book.....W}. During the drag period, baryons decouple from photons and freeze at a scale equal to the acoustic horizon at the drag period redshift.  This scale, known as the sound horizon scale, is a standard ruler embedded in the distribution of galaxies. It is a crucial theoretical prediction of the cosmological model, dependent on the speed of sound in the baryon-photon plasma and the rate of expansion of the early Universe before matter and radiation decoupled. The Sloan Digital Sky Survey (SDSS) \cite{2005ApJ...633..560E} and the 2dF Galaxy Redshift Survey (2dFGRS) \cite{2005MNRAS.362..505C} first detected BAO signal and demonstrated the power of the BAO as a standard ruler for cosmology.  Subsequent BAO surveys including the Six-degree Field Galaxy Survey (6dFGS) \cite{2011MNRAS.416.3017B}, the Baryon Oscillation Spectroscopic Survey (BOSS) \cite{2017MNRAS.470.2617A}, the extended Baryon Oscillation Spectroscopic Survey (eBOSS) \cite{2021PhRvD.103h3533A}, and the WiggleZ Survey \cite{2012MNRAS.425..405B} aimed to achieve more precise cosmic distance measurements at a percentage level. These measurements could offer a highly statistically significant study of the current tension problems  (please see e.g. \cite{2021CQGra..38o3001D,2022JHEAp..34...49A,2022NewAR..9501659P} and  references therein for a more comprehensive discussion).
Currently, the Dark Energy Spectroscopic Instrument (DESI) collaboration has presented Data Release 1 (DR1) of BAO measurements and showed more than 2$\sigma$ evidence for the dynamical dark energy \cite{2024arXiv240403000D,2024arXiv240403001D,2024arXiv240403002D}.
However, before using the BAO as a standard ruler for cosmology and as a powerful cosmological probe,
one needs to know the comoving length of this ruler, i.e., the sound horizon $r_d$ at the radiation drag epoch.

The sound horizon $r_d$ is usually calibrated at $z\approx 1100$ relying on the CMB observations.  Since the length of the scale is not known, BAO can only give a relative measurement of the expansion history. This is similar to the type of Ia supernovae (SNe Ia) acting as cosmological standard candles. If the value of absolute magnitude is not known, then SNe Ia can only provide relative distances. This implies that the Hubble constant $H_0$ and the sound horizon $r_d$ are closely related, that there is strong degeneracy between them, and that they link late and early cosmology. Planck cosmic microwave background (CMB) anisotropy (both temperature and polarization) data reported the $r_d=147\pm0.30$ Mpc with assuming cosmological constant plus cold dark matter  ($\Lambda$CDM) model \cite{2020AA...641A...6P}. An alternative approach is to combine BAO measurements with other low-redshift observations. Combining standard clocks and the local $H_0$ measurement to the SNe and BAO, the work \cite{2014PhRvL.113x1302H} obtained the $r_d=142.8\pm3.7$ Mpc and consisted of the results derived from Planck data. Similarly, considering the same data type, the work \cite{2016JCAP...10..019B} got $r_d=136.80\pm4.0$ Mpc using the spline interpolation method for reconstruction of expansion history, when curvature $\Omega_K$ as a free parameter. Subsequently, under the assumption that the universe is flat,  \citet{2017MNRAS.467..731V} inferred sound horizon $r_d=143.9\pm3.1$ Mpc.
See Refs. \cite{2019MNRAS.486.2184M,2017JCAP...01..015L,
2018PhRvD..98h3526S,2020MNRAS.495.2630C,2021PhRvD.103d3513Z,2024arXiv240607493G} for more works about the sound horizon.
It is important to emphasize here that most of the work to determine sound horizon is cosmological model dependent. If one considers the data from standard clocks to reconstruct the expansion history (eliminating the assumptions of the cosmological background), then one needs to have an assumption about the curvature of the universe or use it as a free parameter. Considering the current critical situation for the measurements of the Hubble constant in astronomical observations, in particular the assumptions of cosmological models, it is very necessary to realize the calibration or determination of $r_d$ in the cosmological model-independent way.

As one of the most ubiquitous phenomena in astronomy, strong gravitational lensing by elliptical galaxies directly provides absolute distance, and is a powerful tool to study the velocity dispersion function of early-type galaxies \citep{2008MNRAS.384..843M,2021MNRAS.503.1319G,2007ApJ...658L..71C}, the distribution of dark matter \citep{2022A&A...659L...5C,1993ApJ...407...33M,2009ApJ...706.1078N}, and cosmological parameters \citep{2014ApJ...788L..35S,2017MNRAS.465.4914B,2019MNRAS.488.3745C}. In particular, gravitational lensing systems with time-delay measurements between multiple images provide a valuable opportunity for the determination of $H_0$.
In representative work of \citet{2020MNRAS.498.1420W}, the $H_0$ Lenses in COSMOGRAIL's Wellspring (H0LiCOW) collaboration combined the six gravitationally lensed quasars with well-measured time delays to constrain the $H_0$. The full dataset consists of six lenses, five of which were analyzed blindly, and four
of which have both time-delay distance $D_{\Delta t}$ and the angular diameter distance to the lens $D_d$ measurements. Current state-of-the-art lensing programs for time-delay with lensed quasars have great progress, such as TDCOSMO collaboration$\footnote{http://tdcosmo.org}$ \cite{2020A&A...639A.101M,2020A&A...642A.193M} (formed by members of H0LiCOW \cite{2020MNRAS.498.1420W}, COSMOGRAIL \cite{2005A&A...436...25E}, STRIDES \cite{2018MNRAS.481.1041T}, and SHARP. Recently, TDCOSMO \cite{2020A&A...643A.165B} based on a new hierarchical Bayesian approach with the original seven lenses, six of which are from H0LiCOW, and reported $H_0=67.4^{+4.1}_{-3.5}$ $\Mpc$. Previously, there has been some work considering the use of time-delay gravitational lensing to calibrate the sound horizon \cite{2019ApJ...874....4A,2019MNRAS.486.5046W,2019A&A...632A..91A}. However, these works were either cosmological model dependent or introduced data like standard clock. Using these data together, it is not possible to disentangle and determine the contribution of these different probes to calibrating the sound horizon in the BAO.

Inspired by the above, this work will use time-delay strong gravitational lensing systems combined with the recent BAO measurements from DES, BOSS/eBOSS, and DESI, to calibrate the sound horizon $r_d$ in a model-independent method.
The combination of these two probes has a number of advantages: first, it is independent of the cosmological model and is independent of the early universe; second, it is independent of cosmological parameters such as the Hubble constant, dark energy, (and even more no cosmic curvature is involved when using the $D_d$ of the lenses, and also avoids the obstacle of mass-sheet degeneracy in gravitational lensing); and third, it does not need to take into account the Eddington relation with respect to the transformation of distance. Fourth, only two types of data are considered, enabling a clear understanding of the contributions of each data.

This paper is organized as follows: in Section 2, we present the data used in this work and the methodology of calibration of the sound horizon. In Section 3, we give our results and discussion. Finally, the main conclusions are summarized in Section 4.

\section{Data and Methodology }\label{sec:data}
\subsection{BAO angular scale measurements}
\begin{figure}
{\includegraphics[width=1\linewidth]{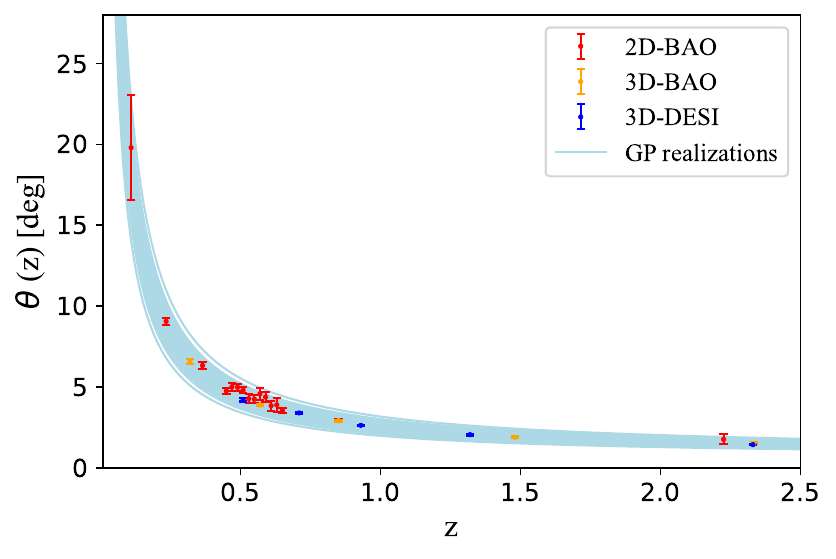}}
\caption{ The points and error bars come from observed BAO angular scale measurements.  The light blue lines are the GPR reconstruction and consist of 1000 curves.
}\label{fig:sn}
\end{figure}

The clustering of matter created by BAO provides a ``standard scale" of length in cosmology. The length of this standard scale (roughly 150 Mpc at present) can be measured by astronomical surveys looking at the large-scale structure of matter, thus constraining the cosmological parameters (especially the density of baryonic matter), and further understanding the nature of the dark energy that causes the accelerated expansion of the Universe.
However, when using BAO for cosmological studies, it is important to know the length of this standard ruler.
During the drag epoch, baryons decoupled from photons and ``froze in" at a scale equal to the sound horizon at the drag epoch redshift $z_d$, i.e., $r_s\equiv r_d(z_d)$, if $r_s$ is interpreted as the sound horizon at radiation drag, then
\begin{equation}
r_d=\int^{\infty}_{z_d}dzc_s(z)/H(z),\label{eq1}
\end{equation}
where $z_d$ being the redshift of the drag epoch, $c_s(z)$ is the sound speed, and $H(z)$ is the Hubble parameter.
The angular BAO scale $\theta_{BAO}$ can be written as
\begin{equation}
	\theta_{BAO} =\frac{r_d}{(1+z)D^A},
\end{equation}
in terms of the angular diameter distance $D^A$.
In this work, we consider the 15 transverse BAO angular scale measurements (denoted as 2D-BAO) summarized in Table 1 of \citet{2020MNRAS.497.2133N}. These values were obtained using public data releases (DR) of the Sloan Digital Sky Survey (SDSS), namely: DR7, DR10, DR11, DR12, and DR12Q (quasars), without assuming any fiducial cosmological model. It is important to note that because these transverse BAO measurements are performed using cosmology-independent methods, their uncertainties are larger than those obtained using the fiducial cosmological method. For anisotropic BAO (denoted as 3D-BAO), we considered two sources of data, a dataset from DES Y6 and BOSS/eBOSS \cite{2017MNRAS.470.2617A,2021PhRvD.103h3533A}, and other from recent DESI DR1 data \cite{2024arXiv240403002D}.
The 3D-BAO  angular scale measurements can be found in Table 1 of \citet{2024arXiv240512142F}.
These BAO angular scale measurements including the 2D-BAO and 3D-BAO are shown in  Fig.~\ref{fig:sn}.

As mentioned above, the joint use of other low redshift observations is necessary if a cosmological model-independent $r_d$ calibration is to be realized. Here we consider observations from the H0LiCOW analysis of six lensed quasars with good lens modeling. The realization of a gravitational lens calibrated sound horizon requires that the BAO be able to provide the corresponding cosmological information both at the redshifts of the sources and lenses.  However, both BAO and gravitational lensing data are very sparse.  Therefore, we consider in this work a cosmological model-independent data reconstruction method, Gaussian Process Regression (GPR) \citep{Holsclaw1,Keeley0,ShafKimLind}, to reconstruct the angular scale measurements of the observed BAO. We generate samples of reconstructed angular scale BAO measurements from the posteriors of 2D and 3D BAO datasets. The posterior sampling of GPR  is realized with the code \texttt{GPHist}\footnote{https://github.com/dkirkby/gphist.} \citep{GPHist}. GPR is a completely data-driven reconstruction method and performs in an infinite dimensional function space without overfitting problem  \citep{Keeley0}. GPR works by generating large samples of functions $\gamma(z)$ determined by the covariance function. The covariance between these functions can be described by the kernel function. We use here the most general and commonly used squared exponential kernel to parameterize the covariance
\begin{equation}
    \langle \gamma(z_1)\gamma(z_2) \rangle = \sigma_f^2 \, \exp\{-[s(z_1)-s(z_2)]^2/(2\ell^2)\},
\end{equation}
where $\sigma_f$ and $\ell$ are hyperparameters and are marginalized over.  The $\gamma(z)$ is a random function inferred from the distribution defined by the covariance, and we adopt $\gamma(z) = \ln(\theta_{BAO}(z))$ to generate more angular scale measurements by using BAO dataset.
The 1000 reconstructed curves of angular scale measurements from the BAO dataset are shown in Fig.~\ref{fig:sn}. It shows the shape of the angular scale-redshift relation of BAO data very well.
It should be noted that the redshift of the BAO dataset well covers the range of H0LiCOW lensing system redshifts, we needn't extrapolate the redshift range of the reconstructed BAO dataset.

\subsection{Strong gravitational lensing systems from H0LiCOW}

Let us briefly outline the standard procedure for $D_{\Delta t}$ and $D_{d}$ measurements used in the H0LiCOW procedure. The distances of a gravitational lens involve only the angular diameter distances.
For a given strong lensing system, quasars act as a background source at redshift $z_{s}$, which is lensed by foreground elliptical galaxies (at redshift $z_{d}$), and multiple bright images of active galactic nuclei (AGN) are formed along with arcs of their host galaxies.
The subscripts $d$ and $s$ stand for the lens galaxy and the source, respectively.
The lensing time delay between any two images is determined by the geometry of the universe and the gravity field of the lensing galaxy \citep{1964PhRvL..13..789S,1964MNRAS.128..307R}
\begin{equation}
\Delta t_{ AB} = D_{ \Delta t}\left[\phi(\theta_{ A},\beta)-\phi(\theta_{ B},\beta)\right]=D_{  \Delta t}\Delta\phi_{ AB}(\xi_{ lens}),
\end{equation}
where $\phi(\theta,\beta)=\left[{(\theta-\beta)^2}/{2}-\psi(\theta)\right]$ is the Fermat potential at images, $\beta$ is the source position, $\psi$ is lensing potential obeying the Poisson equation $\nabla^2\psi=2\kappa$, where $\kappa$ is the surface mass density of the lens in units of critical density $\Sigma_{ crit}=D_{ s}/(4\pi D_{d}D_{ds})$, and $\xi_{lens}$ denotes the lens model parameters. The cosmological background is reflected in the ``time delay distance"
\begin{equation}
D_{\Delta t}=(1+z_{ d})\frac{D_{d}D_{ s}}{D_{ ds}}=\frac{\Delta t_{ AB}}{\Delta\phi_{ AB}(\xi_{ lens})}.
\end{equation}
The variability of the AGN light curve can be monitored to measure the time delay between multiple images.
The key point here is that the Fermat potential difference $\Delta\phi_{ AB}(\xi_{ lens})$ can be reconstructed by high-resolution lensing imaging from space telescopes.

On the other hand, assuming an explicit model for the lens, such as the simplest singular isothermal sphere (SIS) model (not limited to the SIS model, and denoted as lens model parameter $\xi_{{lens}}$), plus observations on stellar kinematics, such as the lens galaxy such as the light profile $\xi_{{light}}$, the line of sight (LOS) projected stellar velocity dispersion of the lens galaxy $\sigma_v$, the anisotropy distribution of the stellar
orbits  $\beta_{{ani}}$, one can yield the absolute distance measure of $D_{d}$ at the lens \citep{2019MNRAS.484.4726B}
\begin{eqnarray}
D_{ d}=\frac{1}{1+z_{ d}}\frac{c\Delta t_{ AB}}{\Delta \phi_{ AB}(\xi_{{lens}})}\frac{c^2J(\xi_{{lens}},\xi_{{light}},\beta_{{ani}})}{\sigma_v^2}\,,
\label{eq:ddp}
\end{eqnarray}
where the function $J$ captures all model components calculated from the lensing image and the photometrically weighted projected velocity dispersion (from spectroscopy).
For more details on the modelling problem for the function $J$, see Section 4.6 of \citep{2019MNRAS.484.4726B}.


The posterior distribution of these lenses (namely RXJ1131-1231 \citep{Suyu13,2014ApJ...788L..35S}, PG1115+080 \citep{Chen19}, B1608+656\footnote{Except for this len, other lenses were analyzed blindly with respect to the cosmological parameters.} \citep{2010ApJ...711..201S,Jee19}, J1206+4332 \citep{2019MNRAS.484.4726B}, WFI2033-4723 \citep{Rusu20}, HE0435-1223 \citep{Wong17,Chen19}) including the time delay distances and the angular diameter distances of the lenses can be found at H0LiCOW website\footnote{http://www.h0licow.org.}.
The redshifts of both lenses and  sources, time delay distances  and  angular diameter distances to lenses for systems are summarized in Table 2 of \citet{2020MNRAS.498.1420W}.

Although  TDCOSMO recently considered different strategies for analyzing these six lenses, compared to the results of H0LiCOW's constraints on the Hubble constant, the results of the latest $H_0$ released by TDCOSMO have a large reduction in precision. This is mainly due to the release of the choice of parameterization of the lens mass profile in H0LiCOW lens modeling, which is the so-called mass-sheet transformation (MST) \cite{1985ApJ...289L...1F}. To counter this increase in uncertainty, the TDCOSMO team obtained stellar kinematics from the Sloan Lens ACS (SLACS) catalogue for constraining the MST. The TDCOSMO IV likelihood products are all publicly available on this repository\footnote{https://github.com/TDCOSMO.} \cite{2020A&A...643A.165B}. However, it is necessary to emphasize that TDCOSMO uses the assumptions of the $\LCDM$ model and the cosmological a priori from SNe Ia in its lensing modeling. Until TDCOSMO realizes a fully cosmological model-independent analysis, we still use the posterior distributions of distances publicly released by H0LiCOW. Although the lensing data are not used the latest, the data from the BAO are the latest, and this is the first work to use gravitational lensing to calibrate the sound horizon of the BAO in a cosmological model-independent way.

\section{Results and Discussion}

Let's emphasize that calibration of BAO sound horizon $r_d$ using observations of the strong lensing system is straightforward, we do not make assumptions about cosmological models and other cosmological parameters.
Combing the Eq. (2) and (5),  the BAO standard ruler $r_d$ can be rewritten as
\begin{equation}
r_d=D_{\Delta t}(\theta_{BAO}(z_d)-\theta_{BAO}(z_s)),
\end{equation}
where $\theta_{BAO}(z_d)$ and $\theta_{BAO}(z_s)$ are  the angular BAO scale at redshifts of source and  lens, and we assume a spatially flat universe and use the standard distance relation to obtain the angular diameter distance between the lens and the source $D_{ds}=D_s-[(1+z_d)/(1+z_s)]D_d$  \citep{Weinberg1972}.
When we combine  the angular diameter distances to lenses with Eq. (2), one can obtain
\begin{equation}
	r_d ={(1+z_d)\theta_{BAO}D_d}. \label{eq5}
\end{equation}
The 1000 curves $\theta_{BAO}$ reconstructed by GPR contain information about its uncertainty. For given a redshift point, we have a distribution of 1000 measurements on the BAO angular scale. Based on these distributions, and combining Eqs. (7) and (8), respectively, we bring them into the skewed log-normal distributions of the time-delay and angular diameter distances provided by gravitational lensing (as published by H0LiCOW) to obtain the Probability Distribution Functions (PDFs) of $r_d$ for the combination of lenses and reconstructed BAO data. Strong lensing systems are in principle uncorrelated, so we obtain the individual distributions of the $r_d$ given by each lens.
\begin{figure}
\begin{center}
\includegraphics[width=1\linewidth]{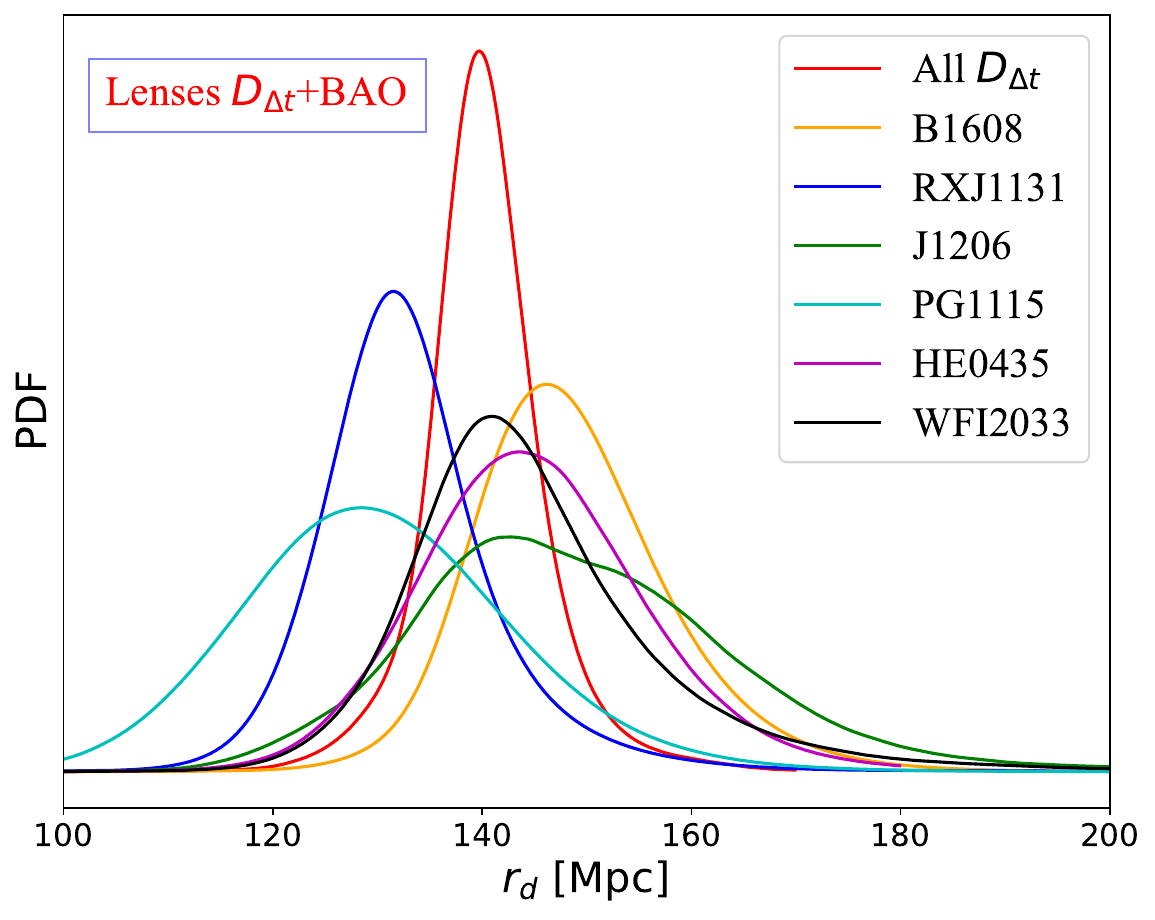}
\end{center}
\caption{The probability density functions of $r_d$ using GPR reconstructed BAO measurements and time delay distances provided by H0LiCOW. }\label{fig2}
\end{figure}
\begin{table*}
\renewcommand\arraystretch{1.5}
\caption{ Summary of the constraints on the sound horizon $r_d$ using recent BAO dataset and H0LiCOW lenses. Comparison with previous works is also provided.}

\begin{center}
\begin{tabular}{l| c| c| c| c |c| c| c}
\hline
\hline
Case &6$D_{\Delta t}$+4$D_{d}$  &6$D_{\Delta t}$ &4$D_{d}$ & Planck18 \cite{2020AA...641A...6P} & Aylor et al. \cite{2019ApJ...874....4A}  &Wojtak et al. \cite{2019MNRAS.486.5046W}& Zhang et al. \cite{2021PhRvD.103d3513Z}
\\
\hline
$r_d$ [Mpc]&$139.7^{+5.2}_{-4.5}$&$139.5^{+5.2}_{-4.4}$&$126.3^{+13.3}_{-10.5}$&$147.05^{+0.3}_{-0.3}$&$139.3^{+4.8}_{-4.8}$
&$137.0^{+4.5}_{-4.5}$&$148.0^{+3.6}_{-3.6}$
\\
\hline
\hline
\end{tabular}
\end{center}\label{rs}
\end{table*}

Using time-delay distances from H0LiCOW, we obtain the final result obtained by combining all lenses is the following:  $r_d=139.5^{+5.2}_{-4.4}$ Mpc (median value with the $16^{th}$ and $84^{th}$ percentiles around it), and are displayed in Fig. \ref{fig2}.  This result is incompatible with the result of Planck 2018 within 1$\sigma$ uncertainty, but consistent within $\sim1.7\sigma$ uncertainty. This is quite reasonable. First of all, H0LiCOW based on these 6 time-delay gravitational lenses gives $H_0=73.3^{+1.7}_{1.8}$ $\Mpc$ in the framework of the $\LCDM$ model.  Planck CMB observation showed their result $H_0=67.4\pm0.5$ $\Mpc$, there is 3.1$\sigma$ tension with H0LiCOW result. Moreover, the Hubble constant and the sound horizon are directly correlated with each other, and there is a strong degeneracy between them and a negative correlation. Our results also highlight the Hubble tension problem to some extent. To understand the relative contributions,
we constrain $r_d$ with each lens in the model-independent manner and also show that Fig. \ref{fig2}.
For each of the individual lens results, they are all compatible with Planck's results within $1\sigma$ uncertainties due to the large uncertainties, except for the lens RXJ1131, its result is $r_d=131.5^{+8.1}_{-6.4}$ Mpc, which is agreement within the $\sim1.7\sigma$ observational uncertainty.  The uncertainty in the GPR reconstructed BAO data is almost equal to the uncertainty in the observed data, and our error budget in the measurement of $r_d$ is clearly dominated by the uncertainty from the observations of the strong lenses, we do not expect any improvement in the precision, as shown by our results. However, such a combination already constrains the low-redshift standard ruler scale $r_d$ at the $\sim3.7\%$ level.
\begin{figure}
\begin{center}
\includegraphics[width=1\linewidth]{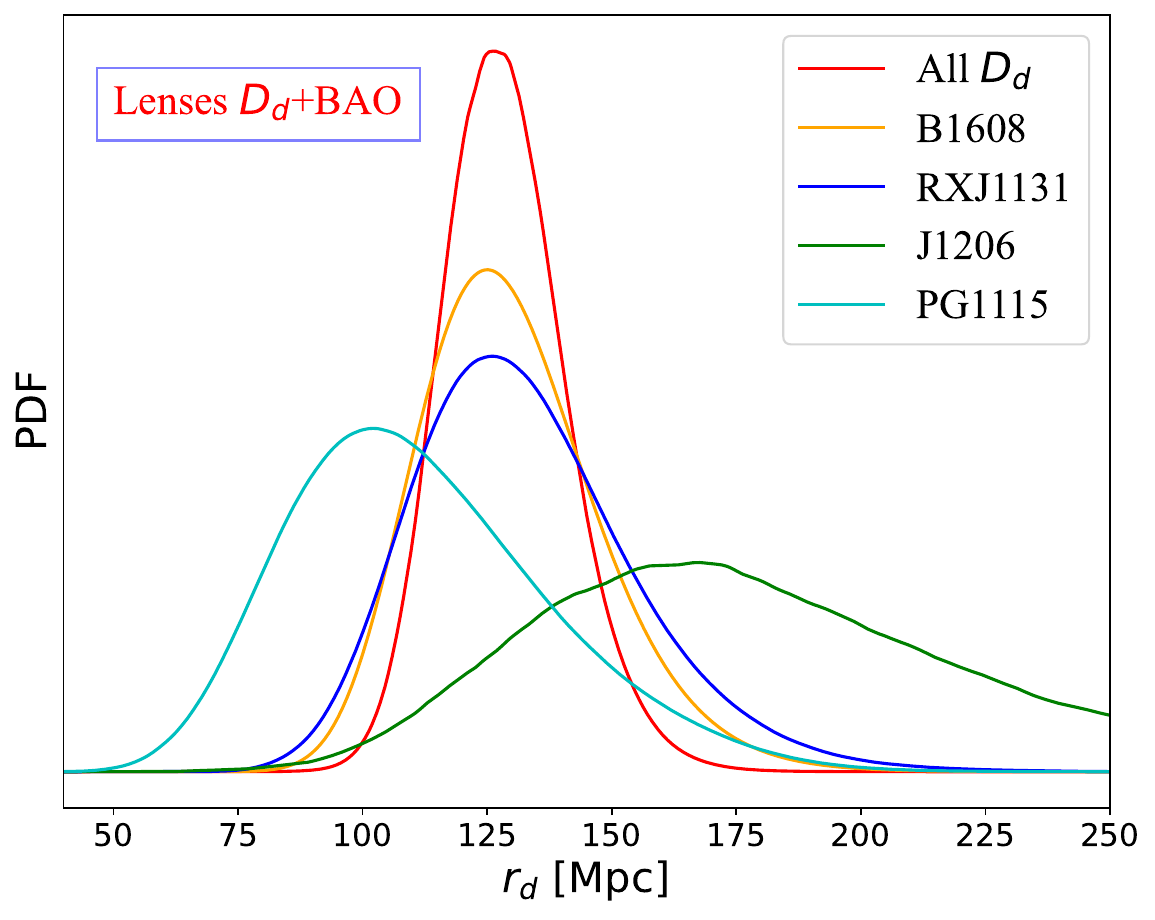}
\end{center}
\caption{The probability density functions of $r_d$ using GPR reconstructed BAO measurements and angular diameter distances provided by H0LiCOW. }\label{fig3}
\end{figure}

Combination of the posterior distributions of angular diameter distances of four H0LiCOW lenses and reconstructed 1000 curves from the BAO dataset, the final PDFs for $r_d$ are reported in Table \ref{rs} and displayed in Fig. \ref{fig3}. Compared with the results of time delay distance, we see that the mean values of $r_d$ for four individual lenses have large changes and large uncertainties.
For a combination of four lenses $D_d$, the corresponding constraint result is $r_d=126.3^{+13.3}_{-10.5}$ Mpc, the constrained precision on $r_d$ is only the $\sim10.5\%$ level, but it is also in agreement with the results of Planck 2018 within $\sim1.5\sigma$ uncertainty. However, it is worth pointing out that, despite the poor precision of the constraints on $r_d$, the method using $D_d$ provided by gravitational lensing has two great benefits. First, the method does not require assumptions about the curvature of the universe, whereas using time-delayed distances requires assumptions to be taken about the curvature. As suggested in work \cite{2019ApJ...874....4A,2019MNRAS.486.5046W}, taking different values for the curvature of the universe might have some impact on the significance of results for measuring $r_d$.
In addition, the method circumvents the mass-sheet degeneracy obstacle in gravitational lensing.

\begin{figure}
\begin{center}
\includegraphics[width=1.0\linewidth]{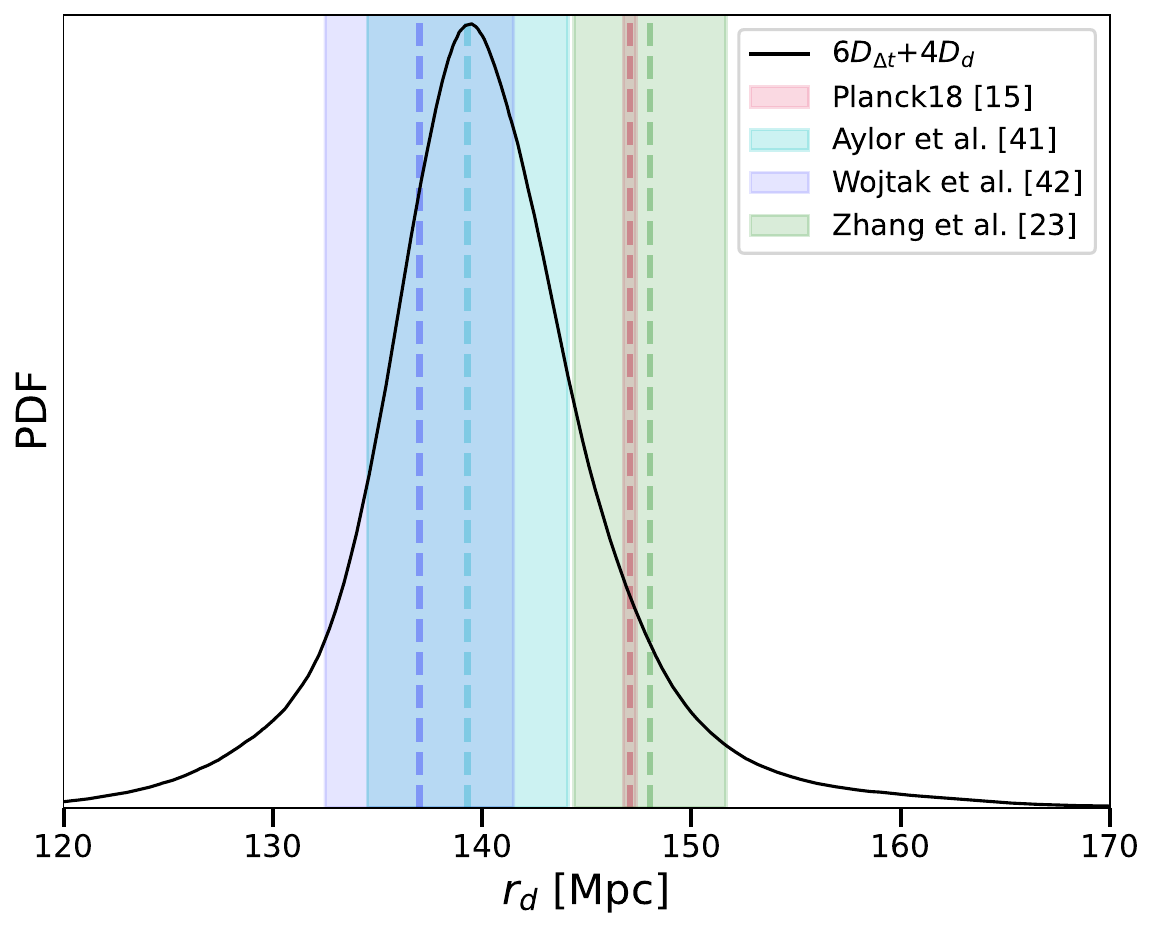}
\end{center}
\caption{The probability density function of $r_d$ using  GPR reconstructed BAO measurements and full data set consisting of 6$D_{\Delta t}$ and 4 $D_d$ measurements from H0LiCOW. Comparison with previous works is also provided.}\label{fig4}
\end{figure}

For the full data set consisting of 6 lenses, 4 of which have both $D_{\Delta t}$ and $D_d$ measurements,
our model-independent constrained result is $r_d=139.7^{+5.2}_{-4.5}$ Mpc and shown in Fig. \ref{fig4}. Compared with results obtained by using 6$D_{\Delta t}$ alone, we see that the mean values of $r_d$ for the full dataset have a few changes (though not significant). As pointed out in the work of \citet{2020A&A...639A.101M}, the addition of $D_d$ (or dispersion measurements) does not play a significant role in the $H_0$ estimation of the H0LiCOW analysis.
However, as we mentioned above, considering different types of data for gravitational lensing has its own advantages. In order to highlight the potential of our method, it is necessary
to compare our results with previous works. Compared our results with previous works, \citet{2019MNRAS.486.5046W} used the $D_d$ of three lenses from the H0LiCOW collaboration, combined relative distances from SNe Ia and BAOs with a prior of $H_0$ from H0LiCOW,  their reported $r_d=137\pm4.5$ Mpc by using cosmography and global fitting method. The work \cite{2021PhRvD.103d3513Z} used the measurements of
BAO, observational $H(z)$ data, GW170817, and SNe Ia to constrain the $H_0$ and $r_d$, and obtained the results of $H_0=68.58\pm1.7$ $\Mpc$ and $r_d=148.0\pm3.6$ Mpc. These works are either cosmological model-dependent or involve more than two or more types of data, which prevents us from clearly understanding the contribution of each type of data. However, it should be stressed that, compared with assuming a specific model, a combination of strong lensing and current astronomical probes such as BAO can reduce possible bias in our work. More importantly, there is no significant increase in uncertainty. As we mentioned earlier (see Introduction), our approach effectively avoids these problems, which suggests that our approach has great potential to provide more precise and accurate measurements of $r_d$ in the future, further precision cosmology research.

\section{Conclusion}

In this work, we propose an improved model-independent method to calibrate sound horizon $r_d$  by using the latest observations of BAO measurement from DES, BOSS/eBOSS, and DESI surveys and gravitationally time-delay lensed quasars from H0LiCOW collaboration.  We adopt the non-parameterized method of GPR to reconstruct observed BAO angular scale measurements. Our approach has several significant advantages.
First, it is independent of the model of the universe and early universe; second, it is independent of cosmological parameters such as the Hubble constant, dark energy, (and, more importantly, does not involve the curvature of the universe when using the $D_d$ of the lenses, and also avoids the obstacle of mass-sheet degeneracy in gravitational lensing); third, it avoids the transformation of distances by the Eddington relation. Fourth, it involves only two types of data, allowing a clear understanding of the contribution of data. Such combinations of time-delay strong lensing systems and current astronomical probes such as BAO can reduce possible bias in the $r_d$ estimation of our analysis.

Combining the 6$D_{\Delta t}$ measurements and reconstructed BAO dataset, our model-independent result is $r_d=139.5^{+5.2}_{-4.4}$ Mpc with the precision at $\sim3.7\%$ level.
This result changes to $r_d=126.3^{+13.3}_{-10.5}$ Mpc when combining 4$D_d$ datasets with the reconstructed BAO dataset. Despite the poor precision of the constraints on $r_d$ in this case, the method using $D_d$ provided by gravitational lensing does not require assumptions about the curvature of the universe and avoids the obstacle of mass-sheet degeneracy, whereas using time-delayed distances requires assumptions to be taken about the curvature.
For the full dataset of 6$D_{\Delta t}$ and 4$D_d$ measurements, we obtain the result $r_d=139.7^{+5.2}_{-4.5}$ Mpc, which is in agreement with the results of Planck 2018 within $\sim1.7\sigma$ uncertainty, the addition of $D_d$ (or dispersion measurements) does not play a significant role in our analysis.  Considering the correlation between $r_d$ and $H_0$,  our results also highlight the Hubble tension and may give us a better understanding of the discordance between the datasets or reveal new physics beyond the standard model.

As a final remark, we are also looking forward to other new cosmological probes, such as gravitational waves as standard sirens.  Recently, the work \cite{2024arXiv240607493G} forecast a relative precision of $\sigma_{r_d}/r_d\sim1.5\%$ from forthcoming surveys such as LISA  gravitational wave (GW) standard sirens and DESI or Euclid angular BAO measurements.  Focusing on the available data before the true gravitational wave era, it has been possible to constrain $r_d$ to up to 3.7$\%$ precision using our method. The methodology of this paper also applies to the $r_d$  estimation of joint GW and BAO  measurements. This shows that our method has great potential to provide more precise and accurate measurements of $r_d$ in the future, further precision cosmology research.

\section*{Acknowledgments}
This work was supported by the National Natural Science Foundation of China under Grant No. 12203009, 12122504 and 12035005; the Chutian Scholars Program in Hubei Province (X2023007); and the Hubei Province Foreign Expert Project (2023DJC040).


\begin{thebibliography}{}
\bibitem[Wald(1984)]{1984ucp..book.....W} Wald R. \ 1984, General Relativity, (University of Chicago Press, Chicago, 1984)
\bibitem[Weinberg(1972)]{1972gcpa.book.....W} Weinberg, S.\ 1972, Gravitation and Cosmology: Principles and Applications of the General Theory of Relativity, by Steven Weinberg, pp. 688. ISBN 0-471-92567-5. Wiley-VCH , July 1972., 688
\bibitem[Eisenstein et al.(2005)]{2005ApJ...633..560E} Eisenstein, D.~J., Zehavi, I., Hogg, D.~W., et al.\ 2005, \apj, 633, 560. doi:10.1086/466512
\bibitem[Cole et al.(2005)]{2005MNRAS.362..505C} Cole, S., Percival, W.~J., Peacock, J.~A., et al.\ 2005, \mnras, 362, 505. doi:10.1111/j.1365-2966.2005.09318.x
\bibitem[Beutler et al.(2011)]{2011MNRAS.416.3017B} Beutler, F., Blake, C., Colless, M., et al.\ 2011, \mnras, 416, 3017. doi:10.1111/j.1365-2966.2011.19250.x
\bibitem[Alam et al.(2017)]{2017MNRAS.470.2617A} Alam, S., Ata, M., Bailey, S., et al.\ 2017, \mnras, 470, 2617. doi:10.1093/mnras/stx721
\bibitem[Alam et al.(2021)]{2021PhRvD.103h3533A} Alam, S., Aubert, M., Avila, S., et al.\ 2021, \prd, 103, 083533. doi:10.1103/PhysRevD.103.083533
\bibitem[Blake et al.(2012)]{2012MNRAS.425..405B} Blake, C., Brough, S., Colless, M., et al.\ 2012, \mnras, 425, 405. doi:10.1111/j.1365-2966.2012.21473.x
\bibitem[Di Valentino et al.(2021)]{2021CQGra..38o3001D} Di Valentino, E., Mena, O., Pan, S., et al.\ 2021, Classical and Quantum Gravity, 38, 153001. doi:10.1088/1361-6382/ac086d
\bibitem[Abdalla et al.(2022)]{2022JHEAp..34...49A} Abdalla, E., Abell{\'a}n, G.~F., Aboubrahim, A., et al.\ 2022, Journal of High Energy Astrophysics, 34, 49. doi:10.1016/j.jheap.2022.04.002
\bibitem[Perivolaropoulos \& Skara(2022)]{2022NewAR..9501659P} Perivolaropoulos, L. \& Skara, F.\ 2022, \nar, 95, 101659. doi:10.1016/j.newar.2022.101659
\bibitem[DESI Collaboration et al.(2024a)]{2024arXiv240403000D} DESI Collaboration, Adame, A.~G., Aguilar, J., et al.\ 2024a, arXiv:2404.03000. doi:10.48550/arXiv.2404.03000
\bibitem[DESI Collaboration et al.(2024b)]{2024arXiv240403001D} DESI Collaboration, Adame, A.~G., Aguilar, J., et al.\ 2024b, arXiv:2404.03001. doi:10.48550/arXiv.2404.03001
\bibitem[DESI Collaboration et al.(2024c)]{2024arXiv240403002D} DESI Collaboration, Adame, A.~G., Aguilar, J., et al.\ 2024c, arXiv:2404.03002. doi:10.48550/arXiv.2404.03002
\bibitem[Planck Collaboration et al.(2020)]{2020AA...641A...6P} Planck Collaboration, Aghanim, N., Akrami, Y., et al.\ 2020, \aap, 641, A6. doi:10.1051/0004-6361/201833910
\bibitem[Heavens et al.(2014)]{2014PhRvL.113x1302H} Heavens, A., Jimenez, R., \& Verde, L.\ 2014, \prl, 113, 241302. doi:10.1103/PhysRevLett.113.241302
\bibitem[Bernal et al.(2016)]{2016JCAP...10..019B} Bernal, J.~L., Verde, L., \& Riess, A.~G.\ 2016, \jcap, 2016, 019. doi:10.1088/1475-7516/2016/10/019
\bibitem[Verde et al.(2017)]{2017MNRAS.467..731V} Verde, L., Bernal, J.~L., Heavens, A.~F., et al.\ 2017, \mnras, 467, 731. doi:10.1093/mnras/stx116

\bibitem[Macaulay et al.(2019)]{2019MNRAS.486.2184M} Macaulay, E., Nichol, R.~C., Bacon, D., et al.\ 2019, \mnras, 486, 2184. doi:10.1093/mnras/stz978

\bibitem[L'Huillier \& Shafieloo(2017)]{2017JCAP...01..015L} L'Huillier, B. \& Shafieloo, A.\ 2017, \jcap, 2017, 015. doi:10.1088/1475-7516/2017/01/015
\bibitem[Shafieloo et al.(2018)]{2018PhRvD..98h3526S} Shafieloo, A., L'Huillier, B., \& Starobinsky, A.~A.\ 2018, \prd, 98, 083526. doi:10.1103/PhysRevD.98.083526
\bibitem[Camarena \& Marra(2020)]{2020MNRAS.495.2630C} Camarena, D. \& Marra, V.\ 2020, \mnras, 495, 2630. doi:10.1093/mnras/staa770
\bibitem[Zhang \& Huang(2021)]{2021PhRvD.103d3513Z} Zhang, X. \& Huang, Q.-G.\ 2021, \prd, 103, 043513. doi:10.1103/PhysRevD.103.043513
\bibitem[Giar{\`e} et al.(2024)]{2024arXiv240607493G} Giar{\`e}, W., Betts, J., van de Bruck, C., et al.\ 2024, arXiv:2406.07493. doi:10.48550/arXiv.2406.07493
\bibitem[Matsumoto \& Futamase(2008)]{2008MNRAS.384..843M} Matsumoto, A. \& Futamase, T.\ 2008, \mnras, 384, 843. doi:10.1111/j.1365-2966.2007.12769.x
\bibitem[Geng et al.(2021)]{2021MNRAS.503.1319G} Geng, S., Cao, S., Liu, Y., et al.\ 2021, \mnras, 503, 1319. doi:10.1093/mnras/stab519
\bibitem[Chae(2007)]{2007ApJ...658L..71C} Chae, K.-H.\ 2007, \apjl, 658, L71. doi:10.1086/516569
\bibitem[Cao et al.(2022)]{2022A&A...659L...5C} Cao, S., Qi, J., Cao, Z., et al.\ 2022, \aap, 659, L5. doi:10.1051/0004-6361/202142694
\bibitem[Mellier et al.(1993)]{1993ApJ...407...33M} Mellier, Y., Fort, B., \& Kneib, J.-P.\ 1993, \apj, 407, 33. doi:10.1086/172490
\bibitem[Newman et al.(2009)]{2009ApJ...706.1078N} Newman, A.~B., Treu, T., Ellis, R.~S., et al.\ 2009, \apj, 706, 1078. doi:10.1088/0004-637X/706/2/1078
\bibitem[Suyu et al.(2014)]{2014ApJ...788L..35S} Suyu, S.~H., Treu, T., Hilbert, S., et al.\ 2014, \apjl, 788, L35. doi:10.1088/2041-8205/788/2/L35
\bibitem[Bonvin et al.(2017)]{2017MNRAS.465.4914B} Bonvin, V., Courbin, F., Suyu, S.~H., et al.\ 2017, \mnras, 465, 4914. doi:10.1093/mnras/stw3006
\bibitem[Chen et al.(2019)]{2019MNRAS.488.3745C} Chen, Y., Li, R., Shu, Y., et al.\ 2019, \mnras, 488, 3745. doi:10.1093/mnras/stz1902
\bibitem[Wong et al.(2020)]{2020MNRAS.498.1420W} Wong, K.~C., Suyu, S.~H., Chen, G.~C.-F., et al.\ 2020, \mnras, 498, 1420. doi:10.1093/mnras/stz3094
\bibitem[Millon et al.(2020)]{2020A&A...639A.101M} Millon, M., Galan, A., Courbin, F., et al.\ 2020, \aap, 639, A101. doi:10.1051/0004-6361/201937351
\bibitem[Millon et al.(2020)]{2020A&A...642A.193M} Millon, M., Courbin, F., Bonvin, V., et al.\ 2020, \aap, 642, A193. doi:10.1051/0004-6361/202038698
\bibitem[Eigenbrod et al.(2005)]{2005A&A...436...25E} Eigenbrod, A., Courbin, F., Vuissoz, C., et al.\ 2005, \aap, 436, 25. doi:10.1051/0004-6361:20042422
\bibitem[Treu et al.(2018)]{2018MNRAS.481.1041T} Treu, T., Agnello, A., Baumer, M.~A., et al.\ 2018, \mnras, 481, 1041. doi:10.1093/mnras/sty2329
\bibitem[Birrer \& Treu(2021)]{2021A&A...649A..61B} Birrer, S. \& Treu, T.\ 2021, \aap, 649, A61. doi:10.1051/0004-6361/202039179
\bibitem[Birrer et al.(2020)]{2020A&A...643A.165B} Birrer, S., Shajib, A.~J., Galan, A., et al.\ 2020, \aap, 643, A165. doi:10.1051/0004-6361/202038861
\bibitem[Aylor et al.(2019)]{2019ApJ...874....4A} Aylor, K., Joy, M., Knox, L., et al.\ 2019, \apj, 874, 4. doi:10.3847/1538-4357/ab0898
\bibitem[Wojtak \& Agnello(2019)]{2019MNRAS.486.5046W} Wojtak, R. \& Agnello, A.\ 2019, \mnras, 486, 5046. doi:10.1093/mnras/stz1163

\bibitem[Arendse et al.(2019)]{2019A&A...632A..91A} Arendse, N., Agnello, A., \& Wojtak, R.~J.\ 2019, \aap, 632, A91. doi:10.1051/0004-6361/201935972
\bibitem[Nunes et al.(2020)]{2020MNRAS.497.2133N} Nunes, R.~C., Yadav, S.~K., Jesus, J.~F., et al.\ 2020, \mnras, 497, 2133. doi:10.1093/mnras/staa2036
\bibitem[Favale et al.(2024)]{2024arXiv240512142F} Favale, A., G{\'o}mez-Valent, A., \& Migliaccio, M.\ 2024, arXiv:2405.12142. doi:10.48550/arXiv.2405.12142
\bibitem[Holsclaw et al.(2010b)]{Holsclaw1} Holsclaw, T., Alam, U., Sans{\'o}, B., et al.\ 2010b, \prl, 105, 241302. doi:10.1103/PhysRevLett.105.241302
\bibitem[Shafieloo et al.(2012)]{ShafKimLind} Shafieloo, A., Kim, A.~G., \& Linder, E.~V.\ 2012, \prd, 85, 123530. doi:10.1103/PhysRevD.85.123530
\bibitem[Joudaki et al.(2018)]{Keeley0} Joudaki, S., Kaplinghat, M., Keeley, R., et al.\ 2018, \prd, 97, 123501. doi:10.1103/PhysRevD.97.123501
\bibitem[Kirkby \& Keeley(2017)]{GPHist} Kirkby, D., Keeley, R. 2017, doi:10.5281/zenodo.999564
\bibitem[Falco et al.(1985)]{1985ApJ...289L...1F} Falco, E.~E., Gorenstein, M.~V., \& Shapiro, I.~I.\ 1985, \apjl, 289, L1. doi:10.1086/184422
\bibitem[Refsdal(1964)]{1964MNRAS.128..307R} Refsdal, S.\ 1964, \mnras, 128, 307. doi:10.1093/mnras/128.4.307
\bibitem[Shapiro(1964)]{1964PhRvL..13..789S} Shapiro, I.~I.\ 1964, \prl, 13, 789. doi:10.1103/PhysRevLett.13.789
\bibitem[Birrer et al.(2019)]{2019MNRAS.484.4726B} Birrer, S., Treu, T., Rusu, C.~E., et al.\ 2019, \mnras, 484, 4726. doi:10.1093/mnras/stz200
\bibitem[Suyu et al.(2013)]{Suyu13} Suyu, S.~H., Auger, M.~W., Hilbert, S., et al.\ 2013, \apj, 766, 70. doi:10.1088/0004-637X/766/2/70
\bibitem[Chen et al.(2019a)]{Chen19} Chen, G.~C.-F., Fassnacht, C.~D., Suyu, S.~H., et al.\ 2019a, \mnras, 490, 1743. doi:10.1093/mnras/stz2547
\bibitem[Suyu et al.(2010)]{2010ApJ...711..201S} Suyu, S.~H., Marshall, P.~J., Auger, M.~W., et al.\ 2010, \apj, 711, 201. doi:10.1088/0004-637X/711/1/201
\bibitem[Jee et al.(2019)]{Jee19} Jee, I., Suyu, S.~H., Komatsu, E., et al.\ 2019, Science, 365, 1134. doi:10.1126/science.aat7371
\bibitem[Wong et al.(2017)]{Wong17} Wong, K.~C., Suyu, S.~H., Auger, M.~W., et al.\ 2017, \mnras, 465, 4895. doi:10.1093/mnras/stw3077
\bibitem[Rusu et al.(2020)]{Rusu20} Rusu, C.~E., Wong, K.~C., Bonvin, V., et al.\ 2020, \mnras, 498, 1440. doi:10.1093/mnras/stz3451
\bibitem[Weinberg(1972)]{Weinberg1972} Weinberg, S.\ 1972, Gravitation and Cosmology: Principles and Applications of the General Theory of Relativity

\end{thebibliography}
\end{document}